\documentclass[pra,twocolumn,titlepage,nofootinbib,amsmath,showpacs]{revtex4}%
\usepackage{graphicx}%
\usepackage{color}
\begin{document}

\title{The $\alpha^2(Z\alpha)^4m$ contributions to the Lamb shift and the fine structure in light muonic atoms}

\author{Evgeny Yu. Korzinin}
\affiliation{D.~I. Mendeleev Institute for Metrology, St.Petersburg,
190005, Russia}
\author{Vladimir G. Ivanov} \affiliation{Pulkovo
Observatory, St.Petersburg, 196140, Russia}
\author{Savely~G.~Karshenboim}
\email{savely.karshenboim@mpq.mpg.de}
\affiliation{Max-Planck-Institut f\"ur Quantenoptik, Garching,
85748, Germany} \affiliation{Pulkovo Observatory, St.Petersburg,
196140, Russia}

\begin{abstract}
Corrections to energy levels in light muonic atoms are investigated
in order $\alpha^2(Z\alpha)^4m$. We pay attention to corrections which
are specific for muonic atoms and include the electron vacuum
polarization loop. In particular, we calculate relativistic and
relativistic-recoil two-loop electron vacuum polarization
contributions. The results are obtained for the levels with $n=1,2$ and
in particular for the Lamb shift ($2p_{1/2}-2s_{1/2}$) and
fine-structure intervals ($2p_{3/2}-2p_{1/2}$) in muonic hydrogen,
deuterium, and muonic helium ions.
\end{abstract}

\maketitle

\section{Introduction}


Precision studies of light muonic atoms allow a determination of
nuclear structure with an accuracy not accessible otherwise. A recent
result of the CREMA collaboration on two lines for the Lamb shift muonic
hydrogen \cite{Nature}, their current evaluation of the Lamb shift
in muonic deuterium and their project on muonic helium Lamb shift
necessitate a  clarification of the related theory.

The problem becomes of  special importance due to a discrepancy of
the value of the proton radius \cite{Nature} derived from the
results on hydrogen and deuterium spectroscopy (see, e.g.,
\cite{paris}) and from electron-proton scattering (see, e.g.,
\cite{sc1:mami}). The situation is reviewed, e.g., in
\cite{codata2010,my_adp}.


A comprehensive compilation of the present theoretical situation on
muonic-hydrogen Lamb shift can be found in recent overviews
\cite{Antognini2012,Jentschura2011,borie2012,Indelicato,Nature}
(see, also, \cite{pachucki1996,pachucki1999,bor_h,VASH-book}).

A theoretical expression for the Lamb shift in muonic hydrogen
comprises a number of terms of a few clearly distinguishable types.
Indeed, there are pure QED corrections and corrections which involve
proton structure. The QED corrections may be of the same type as
in ordinary hydrogen and those need only a rescaling with a
substitution of the electron mass for the muon one. (Since the muon-proton mass
ratio is about 1/9, while the electron-proton mass ratio is about 1/2000,
one has to remember, indeed, higher importance of the recoil
corrections in muonic hydrogen, as well as various reduced-mass
effects.) A review on the Lamb shift in ordinary hydrogen can be
found in \cite{codata2010,my_rep,VASH-book}

In addition to those rescaled terms, there is a number of specific
muonic-hydrogen contributions, which are summarized in Table
\ref{t:sum}. They come from Feynman diagrams with closed electron
loops.

 \begin{table}[htbp]
 \begin{center}{
 \begin{tabular}{|c|r|r|r|r|}
 \hline
Term & \multicolumn{4}{|c|}{$\Delta E(2p_{1/2}-2s_{1/2})$ \ [meV]}\\[0.8ex]
\cline{2-5}
  &$\mu$H\qquad\quad\quad&$\mu$D\qquad\quad\quad&$\mu{\rm ^3He}$\qquad\quad &$\mu{\rm ^4He}$\qquad\quad\\[0.8ex]
 \hline
eVP1     &205.026\,12\phantom{(1)}  & 227.656\,45\phantom{(1)} & 1642.3954\phantom{(0)$^*$} & 1666.2940\phantom{(0)}  \\[0.8ex]
eVP2     &  1.658\,85\phantom{(1)}   & 1.838\,04\phantom{(1)}  & 13.0843\phantom{(0)$^*$}  & 13.2769\phantom{(0)} \\[0.8ex]
eVP3   &  0.007\,52\phantom{(1)}   & 0.008\,42(7)               & 0.073(3)$^*$\phantom{0}   & 0.074(3)\phantom{0}\\[0.8ex]
LbL     & $-0.000\,89(2)$  & $-0.000\,96(2)$  &  $-0.0134(6)^*$ &  $-0.0136(6)$\\[0.8ex]
\hline
 \end{tabular}}
\caption{Specific contributions to the Lamb shift $\Delta
E(2p_{1/2}-2s_{1/2})$ in light muonic atoms up to the order
$\alpha^5m$: hydrogen, deuterium, helium-3 and helium-4 ions. The
results concern one-loop, two-loop, and three-loop eVP contributions
and well as the contribution of the light-by-light scattering block
(Fig.~\ref{fig:lbl}). The results marked with asterisk are obtained
in this paper.
\label{t:sum}}
 \end{center}
 \end{table}


The results obtained up to date for muonic hydrogen include
contributions of the one-loop, two-loop \cite{pachucki1996} and
three-loop \cite{vp31,vp31g,vp32,vp31e} electronic vacuum
polarization (eVP) as well as various contributions of the
electronic block of the light-by-light scattering (LbL) \cite{LbL,
LbL2}. Except for the one-loop eVP contributions, the results are
available only for the leading terms. For the one-loop contribution
additionally to the leading non-relativistic term
\cite{uehl_cl,uehl_an}, also a relativistic non-recoil
\cite{bor_h,pachucki1996,sgk-uehl,uehl_an-rel} and recoil
\cite{Jentschura2011a,a(Za)4m} terms are known. The results are
summarized in Table~\ref{t:sum}.  The $n$-loop results are complete
in a sense that they include all possible contributions of the
related order with $n'$-eVP potentials ($n'\leq n$) and their
iterations. E.g., the eVP2 result in Table~\ref{t:sum} consists of a
contribution of the K\"allen-Sabry potential and of a
double-iteration term with the Uehling potential.

Most of the results mentioned are calculated in the leading non-relativistic
approximation and thus do not contribute to the fine structure. The only
correction among them, relevant for the fine structure, is the one-loop
relativistic contribution \cite{bor_h,pachucki1996,sgk-uehl,uehl_an-rel}.

The two-loop eVP corrections, as mentioned, are known only in the
leading order, which is $\alpha^2(Z\alpha)^2m$, where $Z$ is the
nuclear charge and $m$ is the muon mass, and here we consider
relativistic corrections to them. They are of the order of
$\alpha^2(Z\alpha)^4m$. In muonic atoms a ratio of the muon and
nuclear mass is small, but not very small and in particular in
muonic hydrogen $m/M \sim 0.1$. That means that any more or less
accurate calculation should also involve recoil corrections. Here we
consider them exactly in $m/M$, which is the ratio of the muon  and
nuclear masses.

While the main purpose of this paper is to calculate two-loop
relativistic and relativistic recoil eVP contributions, we also
analyze all other sources of corrections of order of
$\alpha^2(Z\alpha)^4m$ and $\alpha^2(Z\alpha)^4m^2/M$.

\begin{figure}[thbp]
\vspace{-4pt}
\begin{center}
\resizebox{8. cm}{!}{\includegraphics{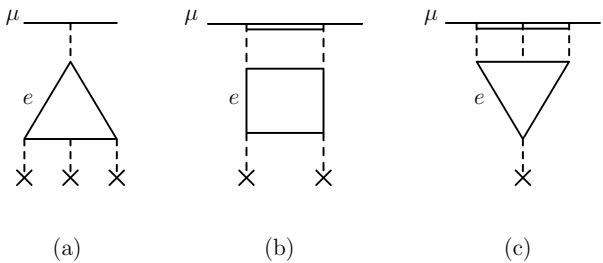}}
\end{center}
\vspace{-7pt} \caption{Leading-order diagrams, which include
the electronic light-by-light scattering block. The horizontal double
line is for the reduced Green function of a muon at the Coulomb field}
\label{fig:lbl}       
\end{figure}

In principle, some of the $\alpha^2(Z\alpha)^4m$ contributions can
appear from the higher-order LbL contributions. The leading LbL
term, presented in Table~\ref{t:sum}, includes the Wichmann-Kroll
contribution (Fig.~\ref{fig:lbl}{\em a}) in order
$\alpha(Z\alpha)^3m$
\cite{EGS,VASH-book,bor_d,bor_he,approx,Blom,LbL, LbL2}, the
virtual-Delbr\"uck-scattering contribution (Fig.~\ref{fig:lbl}{\em
b}) \cite{bor_h,bor_d,bor_he,rmp,scattering,LbL, LbL2} and the third
contribution, which does not have a specific `common' name
(Fig.~\ref{fig:lbl}{\em c}).

Higher-order corrections due to the addition of a radiative correction
to the electron loop or the eVP to either line will
add an extra factor of $\alpha$. However, the LbL term is so
uncertain that such a correction should be below uncertainty.
Besides, it is rather substantially smaller than the two-loop
eVP $\alpha^2(Z\alpha)^4m$ contribution studied in this paper.

\section{Two-loop eVP relativistic recoil contribution}

\begin{figure}[thbp]
\vspace{-4pt}
\begin{center}
\resizebox{7. cm}{!}{ \includegraphics{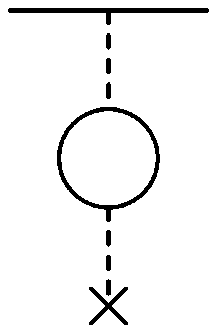}\qquad \qquad
\includegraphics{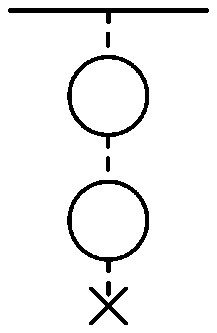}\qquad\qquad
\includegraphics{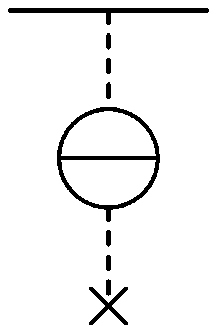}}
\end{center}
\vspace{-7pt} \caption{The eVP potentials are required for a
non-relativistic calculation of the eVP two-loop contribution. They
include the Uehling potential and a reducible and irreducible part of
the K\"allen-Sabry potential \cite{Kallen-Sabry}.}
\label{fig:1NRPT}       
\end{figure}

A calculation of eVP non-relativistic contributions to the energy levels
of a two-body muonic atom can be performed
in terms of the non-relativistic perturbation theory (NRPT). The
only potentials in such a calculation are the Coulomb and eVP
potentials. While the Coulomb problem is considered
non-perturbatively, all the eVP potentials (see
Fig.~\ref{fig:1NRPT}) are considered as a perturbation.
Non-relativistic two-loop \cite{pachucki1996} and
three-loop \cite{vp31,vp31g,vp32,vp31e} eVP terms were found some
time ago  within such an NRPT framework.

In case of the relativistic problem one can apply a Breit-type approach
and also use an NRPT-type calculation, where
in addition to the eVP potentials one has to take into account various
perturbations of the non-relativistic
Hamiltonian that describe the relativistic corrections. That is applicable
for the leading (in $(Z\alpha)$)
relativistic term, but not for higher-order corrections.
Such a leading (in $(Z\alpha)$) term can be found exactly in
$m/M$.

In particular, such a Breit-type approach to the one-loop eVP was
developed in \cite{pachucki1996,pachucki04,Jentschura2011a,a(Za)4m}.

Indeed, for the one-loop eVP correction one can directly calculate
the matrix element over the Dirac-Coulomb wave functions, however,
such a purely relativistic calculation is difficult to generalize to
recoil effects and in particular to the $\alpha(Z\alpha)^4m^3/M^2$ term
as well as the relativistic two-loop terms.

Here we apply the NRPT based on the Breit-type Hamiltonian to the
evaluation of two-loop relativistic corrections and obtain below
the $\alpha^2(Z\alpha)^4m$ term in all orders in $m/M$.

To arrive at an NRPT consideration one has first to consider two
particles which exchange with photons. The NRPT approach eventually
assumes only instantaneous one-photon exchange. Once a Hamiltonian
with instantaneous one-photon exchange is obtained, one can rely on
the perturbation theory
 \begin{eqnarray}\label{basic-VP}
 \Delta E&=&\langle\Psi_{nl}|\delta V|\Psi_{nl}\rangle
 +\langle\Psi_{nl}|\delta V\,G_{nl}^\prime\,\delta V|\Psi_{nl}\rangle\nonumber\\
 &+&\langle\Psi_{nl}|\delta V\,G_{nl}^\prime\,\delta V\,G_{nl}^\prime\,\delta V\,|\Psi_{nl}\rangle  \nonumber\\
 &-&\langle\Psi_{nl}|\delta V|\Psi_{nl}\rangle\times\langle\Psi_{nl}|\delta V\,{G_{nl}^\prime}\,
 {G_{nl}^\prime}\,\delta V|\Psi_{nl}\rangle \,,
 \end{eqnarray}
where $\Psi_{nl}$ is the non-relativistic Coulomb wave function of
the $nl$-state in hydrogenic atom (see, e.g., \cite{III}), $n$ is the principal
quantum number, $l$ is the orbital quantum number. Here,
$G_{nl}^\prime$ stands for the nonrelativistic reduced Coulomb Green function.

The expression is
valid for any central potential. In case of the spin-orbit and
spin-spin interactions the identity needs some corrections. The
interaction of the muon spin and orbit with the nuclear spin is
neglected, since it vanishes after we average over the hyperfine
structure. When necessary, the hyperfine effects can be studied
separately.

As for the spin-orbit interaction, we have to apply the wave
functions the radial part of which is the same as that of $\Psi_{nl}$,
while the angular and spin part is chosen to realize the physical
basis with eigen states of the muon angular momentum and its
projections. Indeed, the matrix elements for the energy do not
depend on the projection.

Strictly speaking, the contributions to the perturbation of the
non-relativistic Hamiltonian, denoted as $\delta V$, are not
necessary potentials, since they may include momentum (see below)
and thus be non-local. That does not change the equations and for
simplicity we still use for then a term  `effective potentials'.

For the one-loop eVP contribution the derivation of the NRPT
equations was done in detail in \cite{a(Za)4m}. A proper choice of
the gauge of the photon propagator $D_{\mu\nu}$ allows to avoid
retardation effects in the $D_{00}$ component of the one-photon
exchange and neglect those effects in the $D_{ij}$ component, since the
retardation effects produce there only corrections in the higher
order in $(Z\alpha)$. Meanwhile the two-photon exchange
contributions lead to $(Z\alpha)^5m^2/M$ terms only. Thus, the
application of the NRPT approach to calculate $\alpha(Z\alpha)^4m$
exactly in $m/M$ is validated.

The evaluation is based on the eVP correction to the photon propagator,
which is proportional to the dispersion integral
(see \cite{a(Za)4m} for details)
\begin{equation}\label{defPmunu}
 D_{\mu\nu}^{\rm VP}(k)\propto \int\limits_0^1{dv}\rho_e(v)\frac{1}{k^2-\lambda^2}\;,
\end{equation}
where the dispersion parameter serves as an effective photon mass
\begin{equation}
\lambda^2=\frac{4m_e^2}{1-v^2}
\end{equation}
and the dispersion function $\rho_e$ depends on the contribution we
are to study. In particular, for the one-loop eVP calculation the
dispersion density is
\begin{equation}\label{rho1}
\rho_1(v)=\left(\frac{\alpha}{\pi}\right)\frac{v^2(1-v^2/3)}{1-v^2}\,.
\end{equation}

The effective potentials at order $\alpha^0$ are determined for a Coulomb-bound two-body system by the standard
Breit equation \cite{breit, breit1}
\begin{eqnarray} \label{VBr}
  V_{\rm  Br}({\bf r})
  &=&-\left(\frac{1}{m^3}+\frac{1}{M^3}\right)\frac{{\bf p}^4}{8}\nonumber\\
  &+&\frac{Z\alpha}{8}\left(\frac{1}{m^2}+\frac{1}{M^2}\right)4\pi \delta^3({\bf r})\nonumber\\
  &+&Z\alpha\left(\frac{1}{4m^2}+\frac{1}{2mM}\right)\frac{{\bf L}\cdot{\mbox{\boldmath $\sigma$}}}{r^3}
    +\frac{Z\alpha}{2mM}4\pi \delta^3({\bf r})\nonumber\\
  &+&\frac{Z\alpha}{2mM}\left[\frac{1}{r^3}{\bf L}^2-{\bf p}^2\frac{1}{r}-\frac{1}{r}{\bf p}^2\right]
 \end{eqnarray}
and considered as a perturbation of the unperturbed problem of the
non-relativistic Schr\"odinger equation with the Coulomb potential
\begin{equation}\label{VC}
 V_C({\bf r})=-\frac{Z\alpha}{r}\;.
 \end{equation}
Here $M$ stands for the nuclear mass, $m$ is for the muon mass
$m_\mu$, ${\bf s}=\mbox{\boldmath $\sigma$}/2$ and ${\bf L}$ are
spin and orbital moments of muon, ${\bf p}$ is the momentum operator
and the relativistic units in which $c=\hbar=1$ are applied. Here
$Z$ is the nuclear change and $M$ is the nuclear mass and the final
expression is valid for the nuclear spin 1/2, assuming that we
average over the nuclear spin (i.e. over the hyperfine structure).

Those in order $\alpha^1$ are \cite{pachucki04}
\begin{eqnarray} \label{Veitia22}
 V_{\rm  Br}^{\rm VP}({\bf r})
 &=&\left(\frac{1}{8m^2}+\frac{1}{8M^2}\right)\nabla^2 V_U\nonumber\\
 &+&\left(\frac{1}{4m^2}+\frac{1}{2mM}\right)\frac{V^{\prime}_U}{r}{\bf L}\cdot{\mbox{\boldmath $\sigma$}}\nonumber\\
 &+&\frac{1}{2mM}\nabla^2\left[V_U-\frac{1}{4}(rV_U)^\prime\right]\nonumber\\
 &+&\frac{1}{2mM}\left[\frac{V^{\prime}_U}{r}{\bf L}^2+\frac{{\bf p}^2}{2}(V_U-rV_U^{\prime})
\right.\nonumber\\
 &&\left. +(V_U-rV_U^{\prime})\frac{{\bf p}^2}{2}\right]\,,
 \end{eqnarray}
where $V_U$ is the Uehling potential
\begin{equation}\label{VU}
V_U({\bf r})=-Z\alpha \int_0^1 dv \, \rho_1(v) \frac{e^{-\lambda r}}{r}\,.
\end{equation}

\begin{figure}[thbp]
\vspace{-4pt}
\begin{center}
\includegraphics{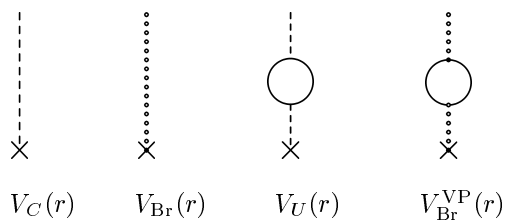}
\end{center}
\vspace{-7pt} \caption{Effective potentials for the NRPT calculation
of the relativistic recoil one-loop eVP contribution.
\label{fig:por-graphs}}
\end{figure}

\begin{figure}[thbp]
\vspace{-4pt}
\begin{center}
\includegraphics[height=3cm]{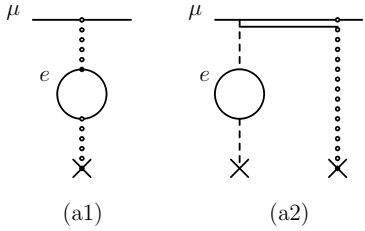}
\end{center}
\vspace{-7pt} \caption{Diagrams for the relativistic recoil one-loop
eVP correction. \label{fig:VP1}}
\end{figure}

Graphically, the related effective potential is presented in
Fig.~\ref{fig:por-graphs} and the diagrams for the calculation of the
relativistic recoil corrections in order $\alpha(Z\alpha)^4m$
(exactly in $m/M$) are depicted in Fig.~\ref{fig:VP1}.

\begin{figure}[ht]
\begin{center}
\includegraphics[height=2.5cm]{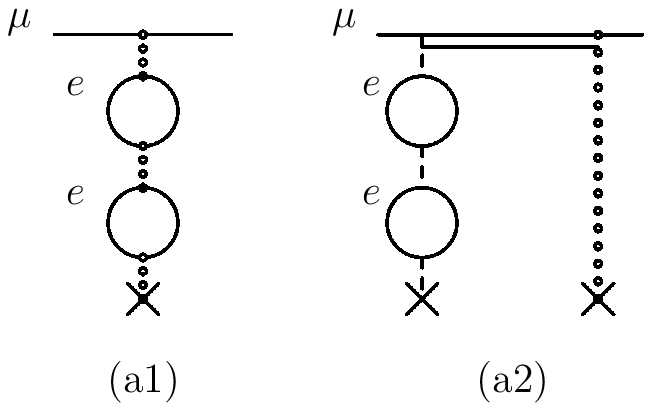}\\ \includegraphics[height=2.5cm]{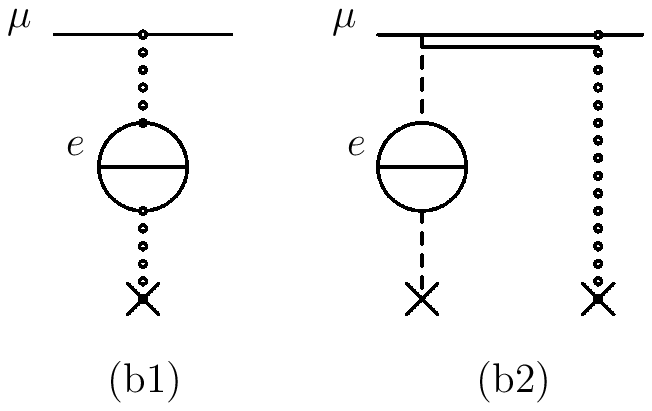}\\ \includegraphics[height=2.5cm]{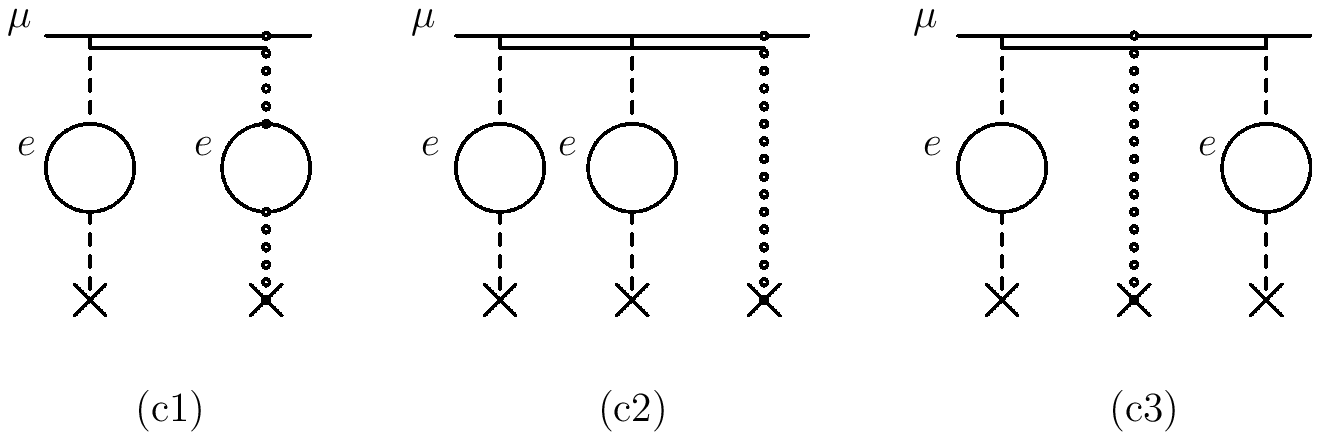}
\end{center}
\caption{Diagrams of the non-relativistic perturbation theory for the
calculation of the relativistic recoil two-loop eVP
contributions\label{fig:all}}
\end{figure}

The relativistic recoil eVP correction of order $\alpha(Z\alpha)^4m$
originates from terms of the first and second
order of NRPT (\ref{basic-VP}) with the effective potential defined
in Eqs. (\ref{VBr}) and (\ref{Veitia22}). To
generalize the result and calculate relativistic recoil two-loop eVP
corrections, we have to calculate terms of the
second (see Fig.~\ref{fig:all} $c1$) and third (see Fig.~\ref{fig:all} $c2$ and $c3$) order with the same potentials
and the first and the second order with the effective two-loop potential (see Fig.~\ref{fig:all} $a$ and $b$). The
latter can be easily obtained from the related one-loop potentials (\ref{Veitia22}) by a substitution of the two-loop eVP
dispersion density for the one-loop one.

The two-loop eVP dispersion function for the reducible part is
\cite{Kallen-Sabry,muhfs}
\begin{eqnarray}\label{rho11}
\rho_{1\cdot1}(v)&=&
 - \frac{1}{9}\left(\frac{\alpha}{\pi}\right)^2\,\frac{v^2(1-v^2/3)}{1-v^2}
 \nonumber\\
 &\times&
 \left\{16-6v^2+3v(3-v^2)\ln\left(\frac{1-v}{1+v}\right)\right\}\;,
\end{eqnarray}
and for the irreducible one it takes the form
\cite{Schwinger,Kallen-Sabry,ro2hfs}
  \begin{eqnarray}\label{rho2}
   \rho_{2}(v)&=&\frac{2}{3}\left(\frac{\alpha}{\pi}\right)^2\frac{v}{1-v^2}\times\biggl\{(3-v^2)(1+v^2)\nonumber\\
   &&\left[{\rm Li}_2\left(-\frac{1-v}{1+v}\right)+2{\rm Li}_2\left(\frac{1-v}{1+v}\right)\right.\nonumber\\
  &+&\left. \ln\left(\frac{1+v}{1-v}\right)\left(\frac{3}{2}\ln\left(\frac{1+v}{2}\right)  -\ln\left(v\right)\right)\right]\nonumber\\
  &+&\left(\frac{11}{16}(3-v^2)(1+v^2)+\frac{1}{4}v^4\right)\ln\left(\frac{1+v}{1-v}\right)\nonumber\\
  &+&\frac{3}{2}v(3-v^2)\ln\left(\frac{1-v^2}{4}\right)-2v(3-v^2)\ln(v)\nonumber\\
  &+&\frac{3}{8}v(5-3v^2)\biggr\}\,,
 \end{eqnarray}
where ${\rm Li}_2$ is the Euler dilogarithm \cite{Lewin}.

The evaluation of the contributions in Figs.~\ref{fig:all}$a$ and $b$
of the reducible and irreducible parts of the two-loop
eVP is similar to the related one-loop eVP contributions
in Fig.~\ref{fig:VP1} and immediately leads to a result. The
non-recoil results  of order ${\alpha}^2(Z\alpha)^4m_rc^2$ for muonic hydrogen are summarized in
Table~\ref{tab:Nonrec-muH}, while the recoil corrections are presented in Table~\ref{tab:1rec-muH}.

\begin{table}[thbp]
\begin{center}
\begin{tabular}{|l|l|l|l|l|}
\hline  
 \parbox{15mm}{Diagram} & \multicolumn{4}{|c|}{$\Delta E(nl_j)$}\\[0.8ex] \cline{2-5}
 &$1s_{1/2}$ & $2s_{1/2}$ &$2p_{1/2}$& $2p_{3/2}$  \\[0.8ex]
 \hline
 \parbox{18mm}{(a1)+2 (a2)}             & $-0.187$         &$-0.0307$          &$-0.000\,957$ &$\phantom{-}0.000\,205$ \\[0.8ex]
 \parbox{18mm}{(b1)+2 (b2)}             & $-0.586$         &$-0.103$           &$-0.0311$     &$-0.003\,75$ \\[0.8ex]
 \parbox{18mm}{2 (c1)\phantom{+(c3)}}   & $\phantom{-}1.16$&$\phantom{-}0.129$ &$-0.000\,995$ &$\phantom{-}0.000\,201$ \\[0.8ex]
 \parbox{18mm}{2 (c2)+(c3)}             & $-1.46$          &$-0.166$           &$-0.000\,758$ &$-0.000\,351$ \\[0.8ex]
  \hline 
  \parbox{15mm}{Total}                  & $-1.07$          &$-0.171$           &$-0.0338$     &$-0.00370$\\[0.8ex]
   \hline
 \end{tabular}
\caption{Relativistic eVP corrections for the low-lying levels in
muonic hydrogen in the external field approximation ($m/M\to0$). The
result is for the Schr\"odinger problem with the reduced mass. The
contributions are labeled as in Fig.~\ref{fig:all}. The units are
$\left({\alpha}/{\pi}\right)^2(Z\alpha)^4m_rc^2$\label{tab:Nonrec-muH}}
 \end{center}
 \end{table}

\begin{table}[thbp]
\begin{center}
\begin{tabular}{|l|l|l|l|l|}
\hline  
 \parbox{15mm}{Diagram} & \multicolumn{4}{|c|}{$\Delta E(nl_j)$}\\[0.8ex] \cline{2-5}
 &$1s_{1/2}$ & $2s_{1/2}$ &$2p_{1/2}$& $2p_{3/2}$  \\[0.8ex]
 \hline
 \parbox{18mm}{(a1)+2 (a2)}           & $\phantom{-}0.454$ &$\phantom{-}0.0648$ &$\phantom{-}0.000\,198$ &$\phantom{-}0.000\,0805$ \\[0.8ex]
 \parbox{18mm}{(b1)+2 (b2)}           & $\phantom{-}0.125$ &$\phantom{-}0.0518$ &$\phantom{-}0.004\,43$  &$\phantom{-}0.001\,67$ \\[0.8ex]
 \parbox{18mm}{2 (c1)\phantom{+(c3)}} & $-3.12$            &$-0.349$            &$-0.000\,413$           &$-0.000\,534$ \\[0.8ex]
 \parbox{18mm}{2 (c2)+(c3)}           & $\phantom{-}3.60$  &$\phantom{-}0.409$  &$\phantom{-}0.000\,796$ &$\phantom{-}0.000\,755$ \\[0.8ex]
  \hline 
  \parbox{15mm}{Total}                & $\phantom{-}1.05$  &$\phantom{-}0.176$  &$\phantom{-}0.005\,01$  &$\phantom{-}0.001\,97$\\[0.8ex]
   \hline
 \end{tabular}
\caption{Relativistic recoil eVP corrections for the low-lying
levels in muonic hydrogen in order ${\alpha}^2(Z\alpha)^4m_rc^2$.
The contributions are labeled as in Fig.~\ref{fig:all}. The units
are
$\left({\alpha}/{\pi}\right)^2(Z\alpha)^4m_rc^2({m_r}/{M})$\label{tab:1rec-muH}}
 \end{center}
 \end{table}

The evaluation of contributions related to Fig.~\ref{fig:all}$c$,
which presents terms of the third order of NRPT (\ref{basic-VP}), is
somewhat more complicated. The related calculation involves
integrations with the radial parts of the reduced Green function of
the non-relativistic Coulomb problem.

We use two representations of the reduced Coulomb Green function
$G_{nl}^\prime$ which allow us to provide a crosscheck of our
calculations. The most fruitful is a representation for the Coulomb
Green function developed in \cite{Hostler-1964}. The expressions we
applied for the radial part of the reduced Coulomb Green functions
are \cite{hameka-1967,Hostler-1969} (see also \cite{pachucki1996})
 \begin{eqnarray}\label{Gr>r<}
\widetilde{G}_{1s}(r,r^\prime)&=&4Z\alpha\,m_r^2\,\exp{\left(\frac{z_>+z_<}{2}\right)}\biggl\{\frac{1}{z_>}+\frac{1}{z_<}\nonumber \\
&~&+\frac{7}{2}-\frac{z_>+z_<}{2}+{\rm Ei}(z_<)-2C \nonumber \\
 &~&-\ln(z_>z_<)-\frac{e^{z_<}}{z_<}\biggr\}\,,
\end{eqnarray}
\begin{eqnarray}
 \widetilde{G}_{2s}(r,r^\prime)&=&Z\alpha\,m_r^2\frac{\exp\left(\frac{z_>+z_<}{2}\right)}{4z_>z_<}\biggl\{8z_<-4z_<^2+8z_> \nonumber\\ \nonumber
 &+& 12 z_>z_< - 26z_>z_<^2 + 2z_>z_<^3 - 4z_>^2 \\ \nonumber
 &-& 26z_>^2z_< + 23z_>^2z_<^2 -z_>^2z_<^3 + 2z_>^3z_< \\ \nonumber
 &-& z_>^3z_<^2 + 4  (z_> - 2)z_>(1 - z_<)e^{z_<}  \\ \nonumber
 &+& 4 (z_> - 2)z_>(z_< - 2)z_< \\
 &\times& \bigl[-2C +{\rm Ei}(z_<) - \ln(z_>z_<)\bigr] \biggr\}\,,
 \end{eqnarray}
\begin{eqnarray}
 \widetilde{G}_{2p}(r,r^\prime)&=&Z\alpha\,m_r^2\frac{\exp\left(\frac{z_>+z_<}{2}\right)}{36(z_>z_<)^2}\biggl\{(24z_<^3 + 36z_>z_<^3  \nonumber\\ \nonumber
 &+& 36z_>^2z_<^3+ 24 z_>^3 + 36z_>^3z_< + 36z_>^3 z_<^2\\ \nonumber
 &+& 49  z_>^3z_<^3 - 3  z_>^3z_<^4-3z_>^4z_<^3  \\ \nonumber
 &-& 12 z_>^3(2+z_< +z_<^2) e^{z_<}+12z_>^3z_<^3 \\
 &\times&  \bigl[-2C +{\rm Ei}(z_<) - \ln(z_>z_<)\bigr]\biggr\}\,,
\end{eqnarray}
where
\begin{eqnarray}
z_>&=&\frac{2Z\alpha\,m_r}{n}{\rm max}(r,r^\prime)\,,\nonumber\\
z_<&=&\frac{2Z\alpha\,m_r}{n}{\rm min}(r,r^\prime)\,,\nonumber
\end{eqnarray}
$C=0.577\,216\ldots$ is the Euler constant, and
\[
{\rm Ei}(x)=\int\limits_{-\infty}^x\frac{e^t}{t}dt
\]
is the exponential integral.

This representation is especially useful in case of contact potentials,
proportional to the $\delta$-function, which sets the smaller radius to
zero (for a general expression for $\widetilde{G}_{nl}(r,0)$ for an
arbitrary state see \cite{vgi-sgk-JETP96}).

The other representation of the reduced Coulomb Green function we used
is the Sturmian one \cite{sturm}. The radial part of the reduced
Coulomb Green function is of the form  \cite{sturm}
 \begin{eqnarray}
 \widetilde{G}_{nl}(r,r^\prime)&=&\frac{n^2}{(Z\alpha)^2m_r}
 \Biggl\{ \sum_{k=l+1\atop k\neq n}^\infty \frac{k}{k-n}R_{kl}(n;r)R_{kl}(n;r^\prime)
 \nonumber\\
 &+& \frac{3}{2}R_{nl}(n;r)R_{nl}(n;r^\prime) \nonumber\\
 &+&  rR^\prime_{nl}(n;r)R_{nl}(n;r^\prime)\nonumber\\
 &+&r^\prime R^\prime_{nl}(n;r^\prime)R_{nl}(n;r)  \Biggr\} \,,\label{gred}
  \end{eqnarray}
and
\[
 R_{kl}(n;r)=\left(\frac{k}{n}\right)^{3/2}R_{kl}\left(\frac{k}{n}r\right)\;,
\]
and $R_{nl}(r)$ stands for the radial part of the standard wave function
of the non-relativistic Coulomb problem (see, e.g., \cite{III}).

An evaluation of the relativistic corrections to the Hamiltonian (\ref{VBr}) and
(\ref{Veitia22}) involves various differentiations and we consider
them in the Appendix. The Laplacian of the Uehling potential is
considered in App.~\ref{s:p2U}, the differentiation of the Green
function in Sturmian representation is discussed in
App.~\ref{s:p2st} and the differentiation procedure applied to the
Green function with $r_{>,<}$ is summarized in App.~\ref{s:p4}. Such
a special treatment of derivatives allows us to simplify the
evaluation. The final results for relativistic non-recoil and recoil
eVP contributions in order $\alpha^2(Z\alpha)^4m$ for the low-lying
states in muonic hydrogen are summarized in
Tables~\ref{tab:Nonrec-muH} and \ref{tab:1rec-muH}.

As for the other light muonic atoms, our results in order
$\alpha^2(Z\alpha)^4m$ and ${\alpha}^2(Z\alpha)^4m_rc^2({m_r}/{M})$
are presented in Tables~\ref{t:4nonrec} and \ref{t:4rec}.

 \begin{table}[htbp]
 \begin{center}
 \begin{tabular}{|l|l|l|l|l|}
 \hline
Atom & \multicolumn{4}{|c|}{$\Delta E(nl_j)$}\\[0.8ex] \cline{2-5}
& $\phantom{-}1s$ & $\phantom{-}2s$ & $\phantom{-}2p_{1/2}$ & $\phantom{-}2p_{3/2}$  \\[0.8ex]
 \hline
$\mu$H    &$-1.07$&$-0.171$&$-0.0338$&$-0.003\,70$   \\[0.8ex]
$\mu$D    &$-1.13$&$-0.180$&$-0.0370$&$-0.004\,15$   \\[0.8ex]
$\mu^3$He &$-2.21$&$-0.347$&$-0.113$&$-0.0163$   \\[0.8ex]
$\mu^4$He &$-2.22$&$-0.350$&$-0.115$&$-0.0165$   \\[0.8ex]
\hline
 \end{tabular}
\caption{Relativistic eVP corrections (Fig.~\ref{fig:all}) for the
low-lying levels in muonic hydrogen in the external field
approximation. The result is for the Schr\"odinger problem with the
reduced mass. The units are
$\left({\alpha}/{\pi}\right)^2(Z\alpha)^4m_rc^2$.\label{t:4nonrec}}
 \end{center}
 \end{table}

 \begin{table}[htbp]
 \begin{center}
 \begin{tabular}{|l|l|l|l|l|}
 \hline
Atom  & \multicolumn{4}{|c|}{$\Delta E(nl_j)$}\\[0.8ex] \cline{2-5}
&\ $1s$ & $\phantom{-}2s$ & $\phantom{-}2p_{1/2}$ & $\phantom{-}2p_{3/2}$  \\[0.8ex]
 \hline
$\mu$H    &$1.05$&$0.176$&$0.005\,01$&$0.001\,97$   \\[0.8ex]
$\mu$D    &$1.13$&$0.191$&$0.004\,53$&$0.002\,78$   \\[0.8ex]
$\mu^3$He &$1.39$&$0.276$&$0.008\,96$&$0.005\,44$   \\[0.8ex]
$\mu^4$He &$1.40$&$0.280$&$0.008\,52$&$0.005\,81$   \\[0.8ex]
\hline
 \end{tabular}
\caption{Relativistic recoil corrections in order
${\alpha}^2(Z\alpha)^4m_rc^2$ (Fig.~\ref{fig:all}) for the low-lying
levels in muonic hydrogen. The units are
$\left({\alpha}/{\pi}\right)^2(Z\alpha)^4m_rc^2({m_r}/{M})$.\label{t:4rec}}
 \end{center}
 \end{table}

We have performed our calculations applying two different representations
of the reduced Coulomb Green function described above. The calculations
were done also with and without the trick with the operator ${\bf p}^4$,
considered in the Appendix (Sect.~\ref{s:p4}). Calculations without the trick
are possible but require more time and are less accurate.
All the results are consistent.

The evaluation based on the Breit-type approach allows to obtain
recoil effects in order $\alpha^2(Z\alpha)^4m$ exactly in $m/M$, and
we have done here such a calculation. However, we have also
performed another evaluation, applying an alternative technique,
which allows terms  linear in $m_r/M$ only. The details will be
published elsewhere \cite{progress}. The results obtained within
these two approaches are consistent. We thus consider our results on
the relativistic recoil two-loop corrections as well established.

\section{Conclusions}

Indeed, the most interesting are not the shifts of energy of
any level by itself, but rather two intervals, namely, the
Lamb-shift ($2p_{1/2}-2s_{1/2}$) and the fine-structure ($2p_{3/2}-2p_{1/2}$)
intervals. The results for the
two-loop eVP contributions (including the previously known leading
term of order $\alpha^2(Z\alpha)^2m$ \cite{pachucki1996})
are summarized in Tables~\ref{t:LS} and
\ref{t:FS}.

The results for different muonic atoms are obtained by the same method,
however, following \cite{Owen,sgk-pach-1995} the so-called Zitterbewegung
term is not included for the muonic deuterium and
helium-4 ion (cf. \cite{Jentschura2011a,a(Za)4m}).

 \begin{table}[htbp]
 \begin{center}
 \begin{tabular}{|l|l|l|l|l|}
 \hline
 & \multicolumn{4}{|c|}{$\Delta E(2p_{1/2}-2s_{1/2})$ \ [meV]}\\[0.8ex] \cline{2-5}
 $\phantom{\alpha^2}$Atom & $\phantom{-1.6}\mu$H & $\phantom{-1.6}\mu$D & $\phantom{-}\mu^3$He & $\phantom{-}\mu^4$He \\[0.8ex]
 \hline
$\alpha^2(Z\alpha)^2m$            & $\phantom{-}1.658\,85$    & $\phantom{-}1.838\,04$   & $13.0843$    & 13.2769   \\[0.8ex]
$\alpha^2(Z\alpha)^4m^*$          & $\phantom{-}0.000\,199$   & $\phantom{-}0.000\,218$  & $\phantom{1}0.005\,82$  & \phantom{1}0.005\,90  \\[0.8ex]
$\alpha^2(Z\alpha)^4m$ & & & &   \\
        (recoil)$^*$                   & $-0.000\,0251$            & $-0.000\,0131$           & $-0.000\,242$& $-0.000\,174$   \\[0.8ex]
\hline
Total                             & $\phantom{-}1.659\,02$    & $\phantom{-}1.838\,24$   & $13.0899$    & 13.2826   \\[0.8ex]
\hline
 \end{tabular}
\caption{The second-order eVP contributions to the Lamb shift
$2p_{1/2}-2s_{1/2}$ in light muonic atoms. The units are meV. The
results marked with asterisk are obtained in this
paper.\label{t:LS}}
 \end{center}
 \end{table}

We note that the recoil effects in order $\alpha^2(Z\alpha)^4m$ are very small for the fine structure. That is because the correction, linear in $m/M$, vanishes (cf. \cite{uehl_an-rel,Mart-FS}) and the remaining term is of order of $(m/M)^2$.

 \begin{table}[htbp]
 \begin{center}
\begin{tabular}{|l|l|l|l|l|}
 \hline
 & \multicolumn{4}{|c|}{$\Delta E(2p_{3/2}-2p_{1/2})$ \ [meV]}\\[0.8ex] \cline{2-5}
$\phantom{\alpha^2}$Atom & $\phantom{-1.}\mu$H & $\phantom{-1.}\mu$D & $\phantom{-}\mu^3$He & $\phantom{-}\mu^4$He \\[0.8ex]
 \hline
$\alpha^2(Z\alpha)^2m$            & $0$    & $0$      & $0$    & 0   \\[0.8ex]
$\alpha^2(Z\alpha)^4m^*$          & $0.000\,0438$     & $0.000\,0502$  & $0.002\,42$  & $0.002\,47$  \\[0.8ex]
$\alpha^2(Z\alpha)^4m$ & & & &   \\
        (recoil)$^*$                 & $-4.5\cdot10^{-7}$ & $-1.4\cdot10^{-7}$   & $-3.2\cdot10^{-6}$& $-1.9\cdot10^{-6}$   \\[0.8ex]
\hline
Total                             & $0.000\,0433$    & $0.000\,0501$   & $0.002\,42$    & $0.002\,47$   \\[0.8ex]
\hline
 \end{tabular}
\caption{The second-order eVP contributions to the fine-structure
interval $2p_{3/2}-2p_{1/2}$ in light muonic atoms. The
units are meV. The results marked with asterisk are obtained in this
paper\label{t:FS}}
 \end{center}
 \end{table}

It is interesting to compare the obtained above two-loop
eVP relativistic contributions with other contributions of the same
order, i.e. of order $\alpha^2(Z\alpha)^4m$. To conclude let us briefly
overview such contributions.

\begin{figure}[htbp]
\begin{center}
\includegraphics{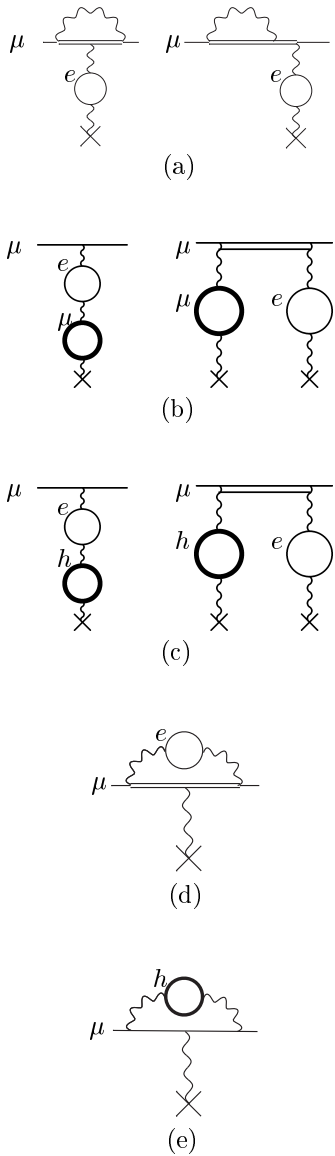}
\end{center}
\caption{Diagrams for various classes of specific corrections,
contributing at order $\alpha^2(Z\alpha)^4m$. $h$ stands for
hadronic vacuum polarization \label{fig:other}}
\end{figure}

Indeed, first of all
there are rescaled contributions of the electronic Lamb shift which
are well known (see, e.g., \cite{codata2010,VASH-book,my_rep}).
Additional specific contributions to the Lamb shift in muonic atoms
in order $\alpha^2(Z\alpha)^4m$ are presented in Fig.~\ref{fig:other}.

As we mention in the introduction, one may also consider
radiative corrections to the block of the light-by-light scattering,
which modify the Wichmann-Kroll potential, and various Uehling corrections
to the leading Wichmann-Kroll contribution. Since the leading Wichmann-Kroll
contributions is very small and the uncertainty of the complete
light-by-light scattering-scattering contribution is not small, we expect
that the $\alpha$ corrections to the leading Wichmann-Kroll contribution
are negligible and below that uncertainty. The related diagrams are not
presented in Fig.~\ref{fig:other}. All the others are. They are split
into several classes.

We remind that
there are non-specific contributions of order $\alpha(Z\alpha)^4m$
which are obtained by the rescaling
\cite{codata2010,VASH-book,my_rep}. Meantime certain corrections to them
with an additional factor of $\alpha$ are already specific. The
typical diagrams are depicted in Fig.~\ref{fig:other}$a$ and $b$.

The $b$-type contributions are due to Uehling corrections to the
contribution of the muon VP. A similar contribution comes from the Uehling
correction to the hadronic vacuum polarization contribution
(see Fig.~\ref{fig:other}$c$).

We note that rescaling should include a substitution of the
mass (electron $\to$ muon), but it keeps the same expression in
terms of $\alpha$ and $Z\alpha$. Technically, that means that we
keep the same expressions for the radiative corrections and in
particular for the anomalous magnetic moment of the muon. The
effects which contribute to the difference in the values of the
anomalous magnetic moments, $a_e$ and $a_\mu$, should be considered
separately. The contributions to $a_\mu-a_e$ appear in order
$\alpha^2$ and most important of them are due to electronic or
hadronic VP (on various contributions to the anomalous magnetic
moment of muon see \cite{a-book1,a-book2}. The characteristic
diagrams are presented in Fig.~\ref{fig:other}$d$ and $e$. Those
diagrams are also responsible for a specific contribution to the
slope of the Dirac form factor in order $\alpha^2$ and thus for the
related contribution to the Lamb shift.

The related contributions are summarized in Tables~\ref{tab:other1}
and \ref{t:other2}. Most of the contributions have been known before.

 \begin{table}[htbp]
 \begin{center}
\begin{tabular}{|c|l|l|l|l|}
 \hline
 & \multicolumn{4}{|c|}{$\Delta E(2p_{1/2}-2s_{1/2})$ \ [$\mu$eV]}\\[0.8ex] \cline{2-5}
 Atom & $\phantom{-.}\mu$H & $\phantom{-.}\mu$D & $\mu^3$He & $\mu^4$He \\[0.8ex] \hline 
 (a)  & $-2.54$       & $-3.06$      & $-62.69$      & $-64.62$   \\[0.8ex]
 (b)  & $\phantom{-}0.128$      & $\phantom{-}0.154^*$    & $\phantom{-}3.83^*$     & $\phantom{-}3.95^*$  \\[0.8ex]
 (c)  & $\phantom{-}0.081(8)^*$ & $\phantom{-}0.097(10)^*$ &\phantom{-} $2.4(2)^*$& \phantom{-}$2.5(2)^*$ \\[0.8ex]
 (d)  & $-1.52$       & $-1.77^*$    & $-29.92^*$    & $-30.73^*$ \\[0.8ex]
 (e)  & $-0.020(2)5^*$      & $-0.024(2)^*$     & $-0.40(4)^*$    & $-0.41(4)^*$ \\[0.8ex]
 eVP2 & $\phantom{-}0.173^*$       & $\phantom{-}0.203^*$    & $\phantom{-}5.58^*$     & $\phantom{-}5.72^*$  \\[0.8ex]
 \hline
 Total& $-3.70(2)$    &  $-4.40(2)$  & $-81.2(2)$    & $-83.6(2)$ \\[0.8ex]
 \hline
 \end{tabular}
\caption{Various $\alpha^2(Z\alpha)^4m_\mu$ contributions the the
Lamb shift ($2p_{1/2}-2s_{1/2}$) in light muonic atoms. The units
are $\mu$eV. Notation follows Fig.~\ref{fig:other}. The eVP2 term is
the correction of order $\alpha^2(Z\alpha)^4m$ of Table~\ref{t:LS}.
The results marked with asterisk are obtained in this paper
\label{tab:other1}}
 \end{center}
 \end{table}

\begin{table}[htbp]
 \begin{center}
\begin{tabular}{|l|l|l|l|l|}
 \hline
 & \multicolumn{4}{|c|}{$\Delta E(2p_{3/2}-2p_{1/2})$ \ [$\mu$eV]}\\[0.8ex] \cline{2-5}
$\phantom{\alpha^2}$Atom & $\phantom{-1.}\mu$H & $\phantom{-1.}\mu$D & $\phantom{-}\mu^3$He & $\phantom{-}\mu^4$He \\[0.8ex]
 \hline
(a)            & $0.0105$         & $0.0127$         & $0.606$    & $0.624$   \\[0.8ex]
(d)$^*$        & $0.0893$         & $0.0991$         & $1.64$     & $1.67$  \\[0.8ex]
(e)$^*$        & $0.0010(1)$     &  $0.0012(1)$      & $0.019(2)$ & $0.020(2)$  \\[0.8ex]
eVP2$^*$       & $0.0433$         & $0.0501$         & $2.42$     & $2.47$ \\[0.8ex]
\hline
 Total         & $0.144$          & $0.164$          & $4.69$     & $4.78$   \\[0.8ex]
\hline
 \end{tabular}
\caption{Various $\alpha^2(Z\alpha)^4m$ contributions to the
fine-structure interval $2p_{3/2}-2p_{1/2}$ in light muonic atoms.
The units are $\mu$eV. Notation follows Fig.~\ref{fig:other}. The
eVP2 term is the correction of order $\alpha^2(Z\alpha)^4m$ of
Table~\ref{t:FS}.  The results marked with asterisk are obtained in
this paper.\label{t:other2}}
 \end{center}
 \end{table}

The type-$a$ contributions to the Lamb shift and fine structure were
considered in \cite{JentschuraWundt}.  The $b$ contributions were
found in \cite{borie75} (see also \cite{VASH-book}) and
\cite{Indelicato} (see also \cite{vp32,muhfs,Antognini2012}).

The contribution, which involves the hadronic vacuum polarization in
the Coulomb photon, (Fig.~\ref{fig:other}$c$) is calculated in this
paper. In particular, we found
\[
\Delta E(2s, \mu{\rm H})=\frac{\alpha}{\pi}\,3.2248...\,\Delta
E_{\rm hadr}^{(0)}(2s, \mu{\rm H})\;,
\]
where $\Delta E_{\rm hadr}^{(0)}(\mu{\rm H})$ is the leading hadronic contribution, considered in the appendix (Sect.~\ref{s:had}).

The contributions $d$ and $e$ are considered for muonic hydrogen in
\cite{Borie82,Barbieri,pachucki1996,VASH-book}. In particular, there is
a result of \cite{Faustov-hadr} for the $e$ contributions. We have
recalculated it and our result is different from that in
\cite{Faustov-hadr}. The details of our calculations as well as a
comparison with the earlier result is presented in
Appendix~\ref{s:f1had}.

The corrections in Tables~\ref{tab:other1} and \ref{t:other2} are
leading non-relativistic corrections in $Z\alpha$ the corresponding
order. They are calculated by means of the non-relativistic atomic
physics, i.e. the related wave functions and Coulomb Green functions
are non-relativistic. That means that all recoil effects are covered
by the reduced mass. We note, that the internal integration of the
radiative loops for the anomalous magnetic moment and the slope of
the Dirac form factor are relativistic. In principle, additionally
to those diagrams one has to take into account diagrams similar to
those in Fig.~\ref{fig:other}$a$, $d$ and $e$ which are radiative
corrections to the nuclear line. However, they are incorporated into
the proton form factors and should be considered separately.

The complete result of the $\alpha^2(Z\alpha)^4m$ (see Table~\ref{tab:other1}) is comparable with the theoretical and experimental uncertainty for the Lamb shift in muonic hydrogen \cite{Nature} and has to be taken into account.

\section*{Acknowledgments}

This work was supported in part by DFG under grant GZ: HA 1457/7-2
and RFBR under grant \#12-02-31741. A part of the work was done
during a stay of VGI and EYK at  the Max-Planck-Institut f\"ur
Quantenoptik, and they are grateful to it for its warm hospitality.

\appendix

\section{Differentiation of the Uehling potential \label{s:p2U}}

Calculation of the Laplacian of the Uehling and K\"allen-Sabry
potentials involves singularities and may cost certain troubles.

Following \cite{pachucki1996,ma-delta,muhfs}, we apply the identity
\[
\nabla^2 V({\bf r})=\int\limits_0^\infty dv
\rho(v)\left(4\pi\delta^3({\bf r})-\frac{\lambda^2}{r}e^{-\lambda
r}\right)\;.
\]

\section{Differentiation for Sturmian basis functions \label{s:p2st}}

While calculating the third order of the NRPT (\ref{basic-VP}) we
have to deal with $\Psi_{nlm} {\bf p}^4 \widetilde{G}$,
$\widetilde{G} {\bf p}^4 \widetilde{G}$, $\Psi_{nlm} A {\bf p}^2
\widetilde{G}$, and $\widetilde{G} A {\bf p}^2 \widetilde{G}$, where
$\Psi_{nlm}$ is the hydrogen wave function and $A$ is an operator,
diagonal in coordinate space. All these expressions require only a
calculation of ${\bf p}^2 \widetilde{G}$ in closed form, since for
${\bf p}^4$ we can consider one ${\bf p}^2$ as acting on the right, while
the other as acting on the left.

Differentiation of the Coulomb wave function is obvious
\begin{equation}\label{wvf}
{\bf p}^2\Psi_{nlm}({\bf
r})=2m_r\left(\frac{Z\alpha}{r}+E_n\right)\Psi_{nlm}({\bf r})\,.
 \end{equation}

A Sturmian basis function
\[
\Phi_{klm}(n,{\bf r})=R_{kl}(n;r)Y_{lm}(\Omega)\,,
\]
is a solution of the related Sturm-Liouville problem, differential
equation for which can be rewritten as
\begin{equation}\label{sturm}
{\bf p}^2\Phi_{klm}(n,{\bf
r})=2m_r\left(\frac{k}{n}\,\frac{Z\alpha}{r}+E_n\right)\Phi_{klm}(n,{\bf
r})\,.
\end{equation}
These identities allow to carry out any differentiation of the Green
function, presented in terms of the Sturmian basis (\ref{gred}), required for the calculation in the third order of NRPT.

\section{Differentiation of the Green function (\ref{Gr>r<}) \label{s:p4}}

The representation of the Green function (\ref{Gr>r<})
in terms of $r_>$ and $r_<$
has certain advantages, however its differentiation is somewhat
complicated. To avoid it we used a trick described below. It may be
applied to any representation of the Green function.

First we note that the NRPT expression (\ref{basic-VP}) was
previously applied in a certain order of the expansion. Having in
mind a calculation of operator ${\bf p}^4/m_r^3$, we consider now
not a single term of the required order ($\alpha^2$), but the sum of
all the terms up the second order in $\alpha$. The sum of all the terms, which
include ${\bf p}^4/m_r^3$, is
 \begin{eqnarray}\label{basic-VPp4}
 \Delta E_4&=&\langle\Psi_{nl}|\frac{{\bf p}^4}{m_r^3}|\Psi_{nl}\rangle\nonumber\\
 &+&\langle\Psi_{nl}|\frac{{\bf p}^4}{m_r^3}\,G_{nl}^\prime\,V_U|\Psi_{nl}\rangle
 +\langle\Psi_{nl}|V_U\,G_{nl}^\prime\,\frac{{\bf p}^4}{m_r^3}|\Psi_{nl}\rangle
\nonumber\\
 &+&\langle\Psi_{nl}|\frac{{\bf p}^4}{m_r^3}\,G_{nl}^\prime\,V_U\,G_{nl}^\prime\,V_U|\Psi_{nl}\rangle  \nonumber\\
 &-&\langle\Psi_{nl}|V_U|\Psi_{nl}\rangle\cdot\langle\Psi_{nl}|\frac{{\bf p}^4}{m_r^3}\,{G_{nl}^\prime}\,{G_{nl}^\prime}\,V_U|\Psi_{nl}\rangle
 \nonumber\\
 &+&\langle\Psi_{nl}|V_U\,G_{nl}^\prime\,V_U\,G_{nl}^\prime\,\frac{{\bf p}^4}{m_r^3}|\Psi_{nl}\rangle  \nonumber\\
 &-&\langle\Psi_{nl}|V_U|\Psi_{nl}\rangle\cdot\langle\Psi_{nl}|V_U\,{G_{nl}^\prime}\,{G_{nl}^\prime}\,\frac{{\bf p}^4}{m_r^3}|\Psi_{nl}\rangle
 \nonumber\\
 &+&\langle\Psi_{nl}|V_U\,G_{nl}^\prime\,\frac{{\bf p}^4}{m_r^3}\,G_{nl}^\prime\,V_U\,|\Psi_{nl}\rangle  \nonumber\\
 &-&\langle\Psi_{nl}|\frac{{\bf p}^4}{m_r^3}|\Psi_{nl}\rangle\cdot\langle\Psi_{nl}|V_U\,{G_{nl}^\prime}\,{G_{nl}^\prime}\,V_U|\Psi_{nl}\rangle \,.
 \end{eqnarray}

To find $\Delta E_4$ one can consider a solution of the
Coulomb-Uehling problem
\begin{equation}\label{eq:CU}
\left(\frac{{\bf p}^2}{2m_r}+V_C+
V_U\right)|\Psi_{nl}^{CU}\rangle=E_{CU}\,|\Psi_{nl}^{CU}\rangle\,.
\end{equation}
The energy and wave function can be presented in terms of series
\begin{equation}
E_{CU}=E^{(0)}+\alpha E^{(1)}+\alpha^2 E^{(2)}+\ldots
\end{equation}
and
\begin{equation}
\Psi_{nl}^{CU}=\Psi^{(0)}_{nl}+\alpha\Psi^{(1)}_{nl}+\alpha^2\Psi^{(2)}_{nl}+\ldots\;.
\end{equation}
Indeed, we can find $E_{CU}$ and $\Psi_{nl}^{CU}$ only using a perturbation
theory with the related leading terms that are the result of solving a pure
Coulomb problem.

Meantime, we note that $\Delta E_4$ has in
these terms a simple form
\begin{equation}
 \Delta E_4=\langle\Psi_{nl}^{CU}|\frac{{\bf p}^4}{m_r^3}|\Psi_{nl}^{CU}\rangle\,,
\end{equation}
which after applying identity (\ref{eq:CU}) can we re-written as
\begin{equation}
 \Delta
 E_4=4\langle\Psi_{nl}^{CU}|\frac{\left(E_{CU}-V_C-V_U\right)^2}{m_r}|\Psi_{nl}^{CU}\rangle\,.
\end{equation}

To obtain $\Delta E_4$ one still has to apply the perturbative
expressions for $E_{CU}$ and $\Psi_{nl}^{CU}$, however, the further
evaluation does not include any derivatives anymore.

A calculation of contributions of $A {\bf p}^2$ can be done similarly.

\section{Leading contribution of the hadronic vacuum polarization\label{s:had}}

\begin{figure}[htbp]
\begin{center}
 \includegraphics{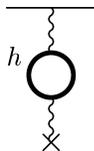}
\end{center}
\caption{The leading hadronic VP contribution to the Lamb
shift\label{fig:had:lead}}
\end{figure}

In the leading order the hadronic vacuum polarization contribution (see
Fig.~\ref{fig:had:lead}) is determined by the value of
polarizability at zero momentum transfer
\[
\frac{{\cal P}_{\rm hadr}(-{\bf k}^2)}{-{\bf k}^4}\biggl|_{{\bf k}^2=0}=-\int_{(2m_\pi)^2}^\infty ds\frac{\rho_{\rm hadr}(s)}{s}\,,
\]
where the dispersion density function can be directly obtained from
experiment by measuring, e.g., the cross section of  $e^+e^-$ annihilation into
hadrons. The leading contribution has roughly order
$\alpha(Z\alpha)^4m$, but it is additionally suppressed by a factor $4m_\mu^2/m_\rho^2$. It was calculated previously for a number of
occasions \cite{VASH-book} (see \cite{Friar1999,Faustov-hadr,hadr-link} for details). Here we recalculate
it.

For the calculation we use a model of the dispersion function applied in \cite{sgk-dimu}. Indeed, we have to update
parameters \cite{RPP} for the hadronic resonances. Following \cite{sgk-dimu}, we estimate the uncertainty at the level
of few percents. The result for the leading hadronic vacuum polarization contribution for the $ns$ state
is
\[
\Delta E_{\rm hadr}^{(0)}(nl)  = -0.169(16)\frac{\alpha}{\pi}\frac{(Z\alpha)^4}{n^3}\frac{m_r^3}{m_\mu^2}\delta_{l0}\;,
\]
which corresponds to $0.0106(11)$ meV for the $2p_{1/2}-2s_{1/2}$ splitting in muonic
hydrogen, which is consistent with the previous calculations \cite{Friar1999,Faustov-hadr}.

The dominant contribution, which is roughly two thirds of the total one,
comes from the pion contributions, which is sufficiently described
by the $\rho$-meson (see, e.g. \cite{GS}). The situation is very similar to that for the hadronic contribution to the anomalous magnetic moment of a muon and for the muonium hyperfine splitting.

If necessary, the leading term can be calculated with accuracy comparable with that for the anomalous magnetic moment of muon (see, e.g., \cite{a-book2,amu_h}) or the muonium hyperfine interval (see, e.g., \cite{mu_h}). That should provide uncertainty below 1\%. However, to calculate higher-order effects, related to diagrams in Fig.~\ref{fig:other}$d$ and $e$, the model considered here is sufficient.

\section{Muon electric form factor with insertion of the hadronic vacuum polarization\label{s:f1had}}

The insertion of the hadronic vacuum polarization into the muon
vertex (see Fig.~\ref{fig:other}$e$) on the mass shell affect both Dirac
($F_1$) and Pauli ($F_2$) form factors. Those induce the
contributions to the energy. The former is determined by the slope
of the Dirac form factor $\partial F_1(q^2)/\partial q^2$ at $q=0$,
and the latter is determined by the value $F_2(0)$, which is the
anomalous magnetic moment of the muon.

While we agree with \cite{Faustov-hadr} on the calculation of the $F_2$
contribution, we do not agree on the $F_1$ contribution. Any vacuum
polarization contribution into the slope can be described by
integrating the Dirac form factor with a non-zero photon mass
$\sqrt{s}$ with a dispersion density function. The slope is of the
form (cf. (11.3.25) in \cite{weinberg})
\begin{eqnarray}\label{Four}
\partial_{q^2}F_1(q^2)\biggr|_{q^2=0}&=&\frac{\alpha}{2\pi}\,\frac{1}{m^2}
\int_0^1dz\left(\frac{1-z^3}{3\,D}\right.\nonumber\\&&+\left.\frac{(1-z)^3(1-4z+z^2)}{6\,D^2}\right)\,,
\end{eqnarray}
where
\begin{equation}
D=(1-z)^2+z \frac{s}{m^2}\,.
\end{equation}
This expression does not agree with \cite{Faustov-hadr}. Actually in
each reference of \cite{Faustov-hadr} a different expression for the
slope is presented and our does not agree with any of them.

To check (\ref{Four}) and alternative expressions from
\cite{Faustov-hadr} we performed several tests. First, we reproduced the
well-known infrared logarithm in the Dirac form factor with $s\to
0$. Only one of three expressions in \cite{Faustov-hadr} reproduced
it. Next, we considered a contribution of insertion of the muon VP
into the muon vertex. It is indeed well known and we reproduced the
known result \cite{slope} from (\ref{Four}), but not from the
expressions in \cite{Faustov-hadr}. Our expression (\ref{Four}) is
consistent with (11.3.25) in \cite{weinberg}.

After those checks we calculated the contribution into the slope of
the Dirac form factor from diagrams in Fig.~\ref{fig:other}$e$ using
the model of the hadronic VP density presented in
Appendix~\ref{s:had}. Our result is presented in
Table~\ref{tab:other1}. It disagrees with results published in
\cite{Faustov-hadr} as well as with those  obtained by us from
their expressions for the slope of the Dirac form factor. We believe
we have performed a sufficient number of tests to rely on our results.

\end{document}